\documentclass[preprint,aps,pra,groupedaddress,amsmath,amssymb,showpacs]{revtex4}
\usepackage{graphicx,amsmath,amssymb,epsfig}
\usepackage{dcolumn}
\usepackage[usenames]{color}
\def\be{\begin{equation}}
\def\ee{\end{equation}}
\def\bea{\begin{eqnarray}}
\def\eea{\end{eqnarray}}
\def\bes{\begin{subequations}}
\def\ees{\end{subequations}}

\begin{document}
\title{Slow-light Airy wave packets and their active control
via electromagnetically induced transparency}
\author{Chao Hang}
\affiliation{State Key Laboratory of Precision Spectroscopy
and Department of Physics,
East China Normal University, Shanghai 200062, China}
\author{Guoxiang Huang}
\email[Corresponding author: ]{gxhuang@phy.ecnu.edu.cn}
\affiliation{State Key Laboratory of Precision Spectroscopy and
Department of Physics, East China Normal University, Shanghai
200062, China}
\date{\today}

\begin{abstract}

We propose a scheme to generate (3+1)-dimensional slow-light Airy
wave packets in a resonant $\Lambda$-type three-level atomic gas
via electromagnetically induced transparency. We show that in the
absence of dispersion the Airy wave packets formed by a probe
field consist of two Airy wave packets accelerated in transverse
directions and a longitudinal Gaussian pulse with a constant
propagating velocity lowered to $10^{-5}\,c$ ($c$ is the light
speed in vacuum). We also show that in the presence of dispersion
it is possible to generate another type of slow-light Airy wave
packets consisting of two Airy beams in transverse directions and
an Airy wave packet in the longitudinal direction. In this case,
the longitudinal velocity of the Airy wave packet can be further
reduced during propagation. Additionally, we further show that the
transverse accelerations (or bending) of the both types of
slow-light Airy wave packets can be completely eliminated and the
motional trajectories of them can be actively manipulated and
controlled by using a Stern-Gerlach gradient magnetic field.

\end{abstract}

\pacs{42.25.-p, 42.65.Jx, 42.65.Tg, 42.50.Gy}
\maketitle


\section{Introduction}

In 1979, Berry and Balaze~\cite{Berry} showed that a
quantum-mechanical Airy wave packet of free particle can be
nondispersive but with constant self-acceleration.
Greenberger~\cite{greenb} argued that such wave packet can be used
to represent a free nonrelativistic particle falling in a constant
gravitational field, and hence the phenomenon obtained is related
to Einstein's equivalence principle.

Based on the similarity between Schr\"{o}dinger equation and
Maxwell equation under paraxial approximation, in recent years
much attention has been paid to the study of Airy light beams or
wave packets~\cite{note001} due to their interesting properties
and potential
applications~\cite{Siviloglou1,Siviloglou2,Bandres,Broky,Sztul,bau,Chong,
Abdollahpour,Polynkin1,Polynkin2,Kasparian,ell,
Jia,Chen,Hu,Lotti,green,Kaminer,Li,Dolev,kam,Panagiotopoulos}. In
particular, generation of three-dimensional (3D) linear Airy light
bullets have also been demonstrated in
experiments~\cite{Chong,Abdollahpour}. However, up to now all Airy
beams or wave packets are considered in passive optical
media~\cite{Siviloglou1,Siviloglou2,Bandres,Broky,Sztul,bau,Polynkin1,Polynkin2,
Kasparian,ell,Chong,Abdollahpour,Jia,Chen,Hu,Lotti,green,Kaminer,Li,Dolev,kam,Panagiotopoulos}.
As a consequence, these light wave packets usually travel with a
speed closed to $c$ (i.e. the light speed in vacuum). Moreover, an
active control on Airy light beams or wave packets is hard to
realize because there is no energy-level structure and selection
rule that can be used and manipulated.

Different from previous studies, in this article we propose a
scheme to generate (3+1)D~\cite{note00} slow-light Airy wave
packets in a resonant $\Lambda$-type three-level atomic gas via
electromagnetically induced transparency (EIT). EIT is a quantum
interference effect in multi-level systems induced by a control
field, by which the propagation of a probe field can display many
striking features, including a significant suppression of optical
absorption, a large reduction of group velocity, as well as a
giant enhancement of Kerr nonlinearity~\cite{fle}. Based on the
EIT technique, an active control of probe-field propagation is
easily achievable due to the existence of energy-level structure
and selection rules.

It is known that the dispersion of an EIT medium is very sensitive
to the time duration  $\tau_0$ of probe
field~\cite{fle,Hua,Hang0}. The dispersion is significant
(negligible) if  $\tau_0$ is small (large). We shall show that
when the dispersion is negligible the Airy wave packets form by a
probe field in our EIT system consist of two Airy wave packets
accelerated in transverse directions and a longitudinal Gaussian
pulse with a constant propagating velocity lowered to $10^{-5}\,c$
($c$ is the light speed in vacuum). We shall also show that when
the dispersion is significant it is able to generate another type
of slow-light Airy wave packets consisting of two Airy beams in
transverse directions and an Airy wave packet in the longitudinal
direction. In this case, the longitudinal velocity of the Airy
wave packet can be further reduced during propagation.
Additionally, we shall further show that the transverse
accelerations (or bending) of the both types of slow-light Airy
wave packets can be completely eliminated and the motional
trajectories of them can be actively manipulated and controlled by
using a Stern-Gerlach (SG) gradient magnetic field. The study
presented here opens an avenue for the exploration of
magneto-optical control on Airy beams and wave packets, and the
results obtained may guide interesting experimental findings of
novel Airy light wave packets and have potential applications in
the field of optical information processing and transmission.

The rest of this article is arranged as follows. In Sec. II, the
physical model and equations of motion under study are given. In
Sec. III, an envelope equation governing the evolution of probe
field for the case of negligible dispersion is derived. The
slow-light Airy wave packet solutions and their acceleration
control by the SG gradient magnetic field are also described. In
Sec. IV, the slow-light Airy wave packets and their active control
for the case of significant dispersion ia studied. Finally, the
last section summarizes the main results obtained in this work.

\section{Model}\label{II}

We consider a cold, lifetime-broadened atomic gas with a $\Lambda$-type level
configuration,
\begin{figure}[tbph]
\centering
\includegraphics[scale=0.5]{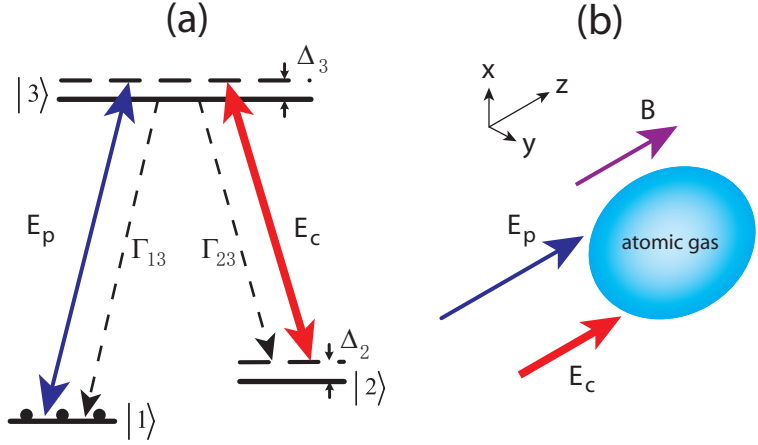}
\caption{{\protect\footnotesize (Color online) (a): Energy-level
diagram and excitation scheme of $\Lambda$-type three-level atoms
interacting with a weak, pulsed probe field $E_{p}$ and a strong,
continuous-wave control field $E_{c}$. $\Delta _{2}$ and $\Delta
_{3}$ are the two- and one-photon detunings, respectively.
$\Gamma_{13}$ ($\Gamma_{23}$) is the decay rate from $|3\rangle$
to $|1\rangle$ ($|3\rangle$ to $|2\rangle$). The initially
populated atoms are indicated by black dots. (b): The coordinate
frame and geometrical arrangement of the system. $B$ is the SG
gradient magnetic field applied to the atomic gas. }} \label{fig1}
\end{figure}
interacting resonantly with a strong, continuous-wave control
field of angular frequency $\omega _{c}$ that drives the
transition $|2\rangle \leftrightarrow |3\rangle $  and a weak,
pulsed probe field (with the time duration $\tau _{0}$ and beam
radius $R$ at the entrance of the medium) of center angular
frequency $\omega _{p}$ that drives the transition $|1\rangle
\leftrightarrow |3\rangle$, respectively; see Fig.~\ref{fig1}(a).
The electric-field vector of the system can be written as
$\mathbf{E}=\mathbf{E}_p+\mathbf{E}_c=\sum_{l=c,p}{\mathbf{e}_{l}\mathcal{E}}_{l}\exp
[i(k_l z-\omega_l t)]+{\rm c.c}.$, where $\mathbf{e}_{c} $ and
$\mathbf{e}_{p}$  ($\mathcal{E}_{c}$ and $\mathcal{E}_{p}$) are,
respectively, the polarization unit vectors (envelopes) of the
control and probe fields. For simplicity, both the probe and
control fields are assumed to propagate along $z$ direction.

Meanwhile, a SG gradient magnetic field with the form
\be \label{SGfield}
{\bf B}(x,y)=\hat{\bf z}B(x,y)=\hat{\bf z}(B_1x+B_2y),
\ee
is applied to the system. Here $\hat{\bf z}$ is the unit vector in the $z$ direction,
$B_1$ and $B_2$ are constants characterizing the magnitudes of the gradient
in $x$ and $y$ directions, respectively. Due to the presence of ${\bf
B}$, Zeeman level shift $\Delta E_{j, \rm Zeeman}=\mu_B g_F^{j}
m_F^{j} B$ occurs for all levels. Here $\mu_B$, $g_F^{j}$, and
$m_F^{j}$ are Bohr magneton, gyromagnetic factor, and magnetic
quantum number of level $|j\rangle$, respectively. The aim of
introducing the SG gradient magnetic field (\ref{SGfield}) is to
provide an external potential to control the accelerating
motion of Airy optical bullets formed in the probe field
(see Sec.~\ref{IIIC} and Sec.~\ref{IV} below). Note that this
technique was also used in a recent study of SG deflection
of slow light and slow-light solitons~\cite{KW,hang}.
A possible geometrical arrangement of
the system is shown in Fig.~\ref{fig1}(b).

Under electric-dipole and rotating-wave approximations, the equations
of motion for the density matrix elements in interaction picture are given by~\cite{boyd}
\begin{subequations}
\label{BLO}
\begin{eqnarray}
& &i\frac{\partial }{\partial t}\sigma _{11}
-i\Gamma _{13}\sigma _{33}+\Omega _{p}^{\ast }\sigma _{31}-\Omega
_{p}\sigma _{31}^{\ast }=0,  \label{1a} \\
& &i\frac{\partial }{\partial t}\sigma _{22}
-i\Gamma _{23}\sigma _{33}+\Omega _{c}^{\ast }\sigma _{32}-\Omega
_{c}\sigma _{32}^{\ast }=0,  \label{1b} \\
& &i\frac{\partial }{\partial t}\sigma _{33}+i\Gamma _{3}\sigma _{33}-\Omega
_{p}^{\ast }\sigma _{31}+\Omega _{p}\sigma _{31}^{\ast }-\Omega _{c}^{\ast
}\sigma _{32}+\Omega _{c}\sigma _{32}^{\ast }=0,  \label{1c} \\
& &\left( i\frac{\partial }{\partial t}+d_{21}\right) \sigma _{21}-\Omega
_{p}\sigma _{32}^{\ast }+\Omega _{c}^{\ast }\sigma _{31}=0,  \label{1d} \\
& &\left( i\frac{\partial }{\partial t}+d_{31}\right) \sigma _{31}-\Omega
_{p}(\sigma _{33}-\sigma _{11})+\Omega _{c}\sigma _{21}=0,  \label{1e} \\
& &\left( i\frac{\partial }{\partial t}+d_{32}\right) \sigma
_{32}-\Omega _{c}(\sigma _{33}-\sigma _{22})+\Omega _{p}\sigma
_{21}^{\ast }=0,
\end{eqnarray}
\end{subequations}%
where the Rabi frequencies of the probe and control fields are
defined, respectively, by
$\Omega _{p}=\mathbf{e}_{p}\cdot
\mathbf{p}_{31}\mathcal{E}_{p}/\hbar $ and $\Omega
_{c}=\mathbf{e}_{c}\cdot \mathbf{p}_{32}\mathcal{E}_{c}/\hbar$,
 with $\mathbf{p}_{jl}$ being the electric dipole
matrix element associated with the transition from states
$|l\rangle$ to $|j\rangle$. In Eq.~(\ref{BLO}), we have also defined
$d_{21}=\Delta_{2}+i\gamma_{21}$,
$d_{31}=\Delta_{3}+i\gamma_{31}$, and
$d_{32}=(\Delta_{3}-\Delta_{2})+i\gamma_{32}$. Here
$\Delta_{2}=(\omega_{p}-\omega_{c}-\omega_{21})+\mu_{21}B$
and $\Delta_{3}=(\omega_{p}-\omega_{31})+\mu_{31}B$ are, respectively, the
two- and one-photon detunings, with
$\mu_{jl}=\mu_B(g_F^{j}m_F^{j}-g_F^{l}m_F^{l})/\hbar$ and
$\omega_{jl}=(E_j-E_l)/\hbar$ ($E_j$ is the eigenenergy of the
state $|j\rangle$). The composite decay rate $\gamma_{jl}$ in $d_{jl}$
is given by
$\gamma_{jl}=(\Gamma_j+\Gamma_l)/2+\gamma_{jl}^{\rm col}$. Here
$\Gamma_j=\sum_{j<l}\Gamma_{jl}$, with $\Gamma _{jl}$ being the
spontaneous emission decay rate from $|l\rangle$ to $|j\rangle$
and $\gamma _{jl}^{\rm col}$ being the dephasing rate reflecting
the loss of phase coherence between $|j\rangle$ and $|l\rangle$
without changing of population~\cite{boyd}.

The equation of motion for $\Omega_p$ can be obtained by
the Maxwell equation, which under the slowly-varying envelope
approximation reads
\begin{equation}\label{MAX}
i\left( \frac{\partial }{\partial z}+\frac{1}{c}\frac{\partial }{\partial t}%
\right) \Omega _{p}+\frac{c}{2\omega _{p}}\left(\frac{\partial
^{2}}{\partial x^{2}}+\frac{\partial ^{2}}{\partial
y^{2}}\right)\Omega _{p}+\kappa _{13}\sigma _{31}=0,
\end{equation}
where $\kappa _{13}=N_{a}\omega
_{p}|\mathbf{p}_{13}|^{2}/(2\varepsilon_{0}c\hbar )$ with
$N_{a}$ being atomic concentration.

The above model can be easily realized in realistic physical
systems. One of candidates is a cold $^{85}$Rb atomic gas with
energy-levels assigned as $|1\rangle=|5^2S_{1/2},F=2,m_F=0\rangle$
($g_F=-1/3$), $|2\rangle=|5^2S_{1/2},F=3,m_F=2\rangle$
($g_F=1/3$), and $|3\rangle=|5^2P_{1/2},F=3,m_F=1\rangle$
($g_F=1/9$). Then the probe field is $\sigma^{+}$-polarized while
the control field is $\sigma^{-}$-polarized. The decay rates are
given by $\Gamma_{13}\approx\Gamma_{23}\approx\pi\times5.75$ MHz,
$\gamma_{13}^{\rm col}\approx\gamma_{23}^{\rm col}\approx1$ kHz,
and $|{\bf p}_{13}|= 2.54\times10^{-27}$ C cm~\cite{Steck}. The
atomic concentration is taken as $N_a=3.67\times10^{10}$
cm$^{-3}$, and hence $\kappa_{13}$ takes the value of $1.0\times
10^{9}$ cm$^{-1}$s$^{-1}$. All calculations given below will be
based on these physical parameters.


\section{Slow-light Airy wave packets in the absence of dispersion}\label{III}

\subsection{Envelope equation}\label{IIIA}

One of the main aims of the present work is to obtain
shape-preserving Airy light wave packets without using any
external potential~\cite{note01}. To this end, we first derive an
envelope equation in the absence of dispersion based on the
Maxwell-Bloch (MB) Eqs.~(\ref{BLO}) and (\ref{MAX}).

We take the following asymptotic expansions
$\sigma_{jk}=\delta_{j1}\delta_{k1}+\epsilon\sigma_{jk}^{(l)}+\epsilon^2\sigma_{jk}^{(2)}$
($j,\,k=1,\,2,\,3$; both $\delta_{j1}$ and $\delta_{k1}$ are
Kronecker delta symbols), $\Omega
_{p}=\epsilon\Omega_{p}^{(1)}+\epsilon^2\Omega_{p}^{(2)}$,
$d_{j1}=d_{j1}^{(0)}+\epsilon d_{j1}^{(1)}$ ($j=2,\,3$), and
$d_{32}=d_{32}^{(0)}+\epsilon d_{32}^{(1)}$, where $\epsilon$ is a
dimensionless small parameter characterizing the amplitude of the
probe field. All quantities on the right hand side of the
expansions are assumed as functions of the multi-scale
variables $z_{2j}=\epsilon^{j} z$, $t_{2j}=\epsilon^{j}
t$ ($j=0,1$), $x_{1}=\epsilon^{1/2} x$, and $y_{1}=\epsilon^{1/2} y$.
Additionally, we assume the gradient of the SG magnetic field is
of $\epsilon^{3/2}$ order, and hence $B(x,y)=\epsilon
(B_1^{(1)}x_1+B_2^{(1)}y_1)$. Thus we have
$d_{21}^{(0)}=(\omega_p-\omega_c-\omega_{21})+i\gamma_{21}$,
$d_{31}^{(0)}=(\omega_p-\omega_{31})+i\gamma_{31}$,
$d_{32}^{(0)}=(\omega_c-\omega_{32})+i\gamma_{32}$,
$d_{21}^{(1)}=\mu_{21}(B_1^{(1)}x_1+B_2^{(1)}y_1)$,
$d_{31}^{(1)}=\mu_{31}(B_1^{(1)}x_1+B_2^{(1)}y_1)$, and
$d_{32}^{(1)}=\mu_{32}(B_1^{(1)}x_1+B_2^{(1)}y_1)$.

Substituting the expansions into the MB Eqs.~(\ref{BLO}) and (\ref{MAX}),
and comparing the coefficients of $\epsilon^l$ ($l=1,2 \cdots$),
we obtain a set of linear but inhomogeneous equations
which can be solved order by order.

At leading order ($l=1$, i.e. the terms of order of $\epsilon$),
we get the solution
\bes \label{ord1} \bea
&&\Omega_{p}=F e^{i[K(\omega)z-\omega t]}, \\
&&\sigma_{j1}=\frac{\delta_{j3}(\omega+d_{21}^{(0)})
-\delta_{j2}\Omega_{c}^{\ast}}{D(\omega)}F e^{i[K(\omega)z-\omega
t]} \quad (j=2,3), \eea \ees
where
$D(\omega)=|\Omega_{c}|^{2}-(\omega+d_{21}^{(0)})(\omega+d_{31}^{(0)})$
and $F$ is a yet to be determined envelope function of the slow
variables $t_{2}$, $x_{1}$, $y_{1}$, and $z_{2}$. The
dependence of $K$ on $\omega$~\cite{note1} obeys the linear
dispersion relation
\be\label{Disp}
K(\omega)=\frac{\omega}{c}+\kappa_{13}\frac{\omega+d_{21}^{(0)}}{D(\omega)}.
\ee

\begin{figure}[tbph]
\centering
\includegraphics[scale=0.75]{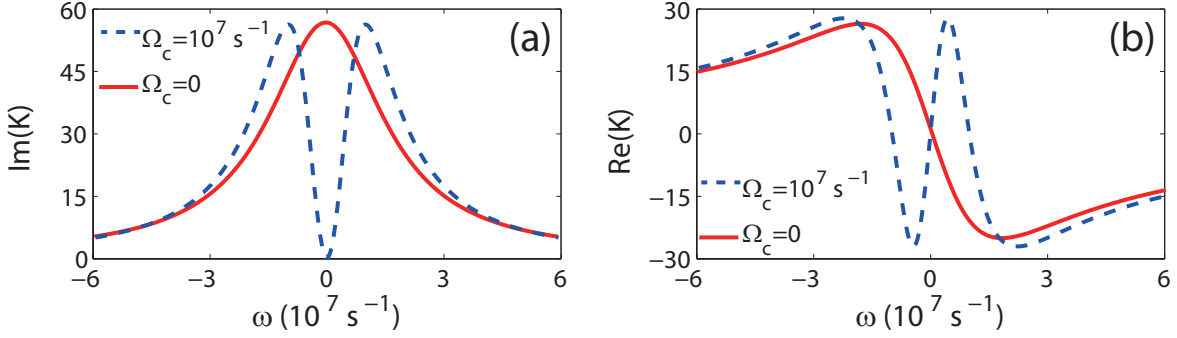}
\caption{{\protect\footnotesize (Color online) Im$(K)$ (a) and
Re$(K)$ (b) as functions of $\omega$, respectively. The dashed and solid lines
in each panel correspond to the presence ($\Omega_{c}=1.0\times
10^{7}$ ${\rm s}^{-1}$) and the absence ($\Omega_{c}=0$) of the
control field, respectively. }} \label{fig2}
\end{figure}

In Fig.~\ref{fig2}(a)
%
%
and Fig.~\ref{fig2}(b) we have plotted the real and imaginary parts of $K$, i.e.
Re$(K)$ and Im$(K)$, as functions of $\omega$ for the exact one-
and two-photon resonances ($\omega_p-\omega_c-\omega_{21}=\omega_p-\omega_{31}=0$). The solid
and dashed lines in the figure correspond, respectively, to the
absence ($\Omega_{c}=0$) and the presence ($\Omega_{c}=1.0\times
10^{7}$ ${\rm s}^{-1}$) of the control field. One sees that when
$\Omega_{c}=0$, the probe field suffers maximum absorption at
$\omega=0$ (the solid line of Fig.~\ref{fig2}(a)\,). However, when
$\Omega_{c}\neq 0$ and satisfies the condition
$|\Omega_c|^2\gg\gamma_{21}\gamma_{31}$, a transparency window is
opened in the probe-field absorption spectrum nearly $\omega=0$
(the dashed line of Fig.~\ref{fig2}(a)\,), and hence the probe field can
propagate in the system with nearly
vanishing absorption, which is a typical character of EIT. In
addition, for the large control field the slope of Re$(K)$ is
drastically changed and steepened around $\omega=0$ (see the
dashed line of Fig.~\ref{fig2}(b)\,) which results in a significant
reduction of the group velocity of the probe field,
$V_g \equiv {\rm Re}[(\partial K/\partial \omega)^{-1}]\approx|\Omega_c|^2/\kappa_{13}$,
and hence slow light (see Eq.~(\ref{groupvelocity}) below).
These interesting characters are due to the quantum destructive
interference (EIT) effect induced by the control field~\cite{fle}.

At next order ($l=2$, i.e. the terms of order of $\epsilon^2$), a
divergence-free condition requires the equation for the envelope
function $F$:
\be \label{ord2}  i\left(\frac{\partial }{\partial
z_{2}}+\frac{1}{V_g}\frac{\partial }{\partial
t_{2}}\right)F+\frac{c}{2\omega _{p}}\left( \frac{\partial ^{2}}{\partial
x_{1}^{2}}+\frac{\partial ^{2}}{\partial y_{1}^{2}}\right)F+
P(x_1,y_1)F=0. \ee
where $V_g \equiv {\rm Re}[(\partial K/\partial \omega)^{-1}]$
is the group velocity of the envelope $F$, and
\be
P(x_1,y_1)=\kappa_{13}\frac{|\Omega_{c}|^2\mu_{21}+(\omega
+d_{21}^{(0)})^2\mu_{31}}{D^{2}(\omega)}(B_1^{(1)}x_1+B_2^{(1)}y_1)
\nonumber
\ee
provides an external potential for $F$,  resulted
from the SG gradient magnetic field (\ref{SGfield}).

Equation (\ref{ord2}) can be written into the dimensionless form
\be\label{EQU1} i\left(\frac{\partial }{\partial
s}+\lambda\frac{\partial }{\partial
\tau}\right)U+\frac{1}{2}\left(\frac{\partial^{2} }{\partial
\xi^{2}}+\frac{\partial^{2}}{\partial \eta^{2}}\right)U
+Q(\xi,\eta)U =0, \ee
with
\be\label{lambda}
\lambda=L_{\rm Diff}/(V_g\tau_0),
\ee
where $s=z/L_{\rm Diff}$ ($L_{\rm Diff}=\omega_p R^2/c$ being the
characteristic diffraction length), $\tau=t/\tau_0$ ($\tau_0$
being the typical duration of the probe filed),
$(\xi,\eta)=(x,y)/R$ ($R$ being the typical transverse radius of
the probe filed), $U=\Omega_{p}/U_0$ ($U_0$ being the typical Rabi
frequency of the probe filed), and $Q(\xi,\eta)=P(x_1,y_1)L_{\rm
Diff}$. We have also assumed that the imaginary part of the
coefficients in the equation is much smaller than the
corresponding real part. This assumption is allowed because of the
existence of the EIT effect induced by the control field (also see
the example given below).

For the convenience of following discussions, we focus on a
particular example under a set of realistic parameters:
$\Omega_c=1.6\times10^7$ s$^{-1}$,
$\omega_p-\omega_c-\omega_{21}=8.0\times10^5$ s$^{-1}$,
$\omega_p-\omega_{31}=-4.0\times10^7$ s$^{-1}$, and
$R=4.78\times10^{-3}$ cm with other parameters being the same with
those used in the last paragraph of Sec.~{\ref{II}. Then we obtain
$K|_{\omega=0}=2.77+i0.14$ cm$^{-1}$ and $\partial
K/\partial\omega|_{\omega=0}=(3.07+i0.31)\times10^{-6}$ cm$^{-1}$
s. We see that the imaginary part of these quantities is indeed
much smaller than their corresponding real part.  Furthermore, the
dispersion length $L_{\rm Diff}=1.81$ cm and the group velocity
\be\label{groupvelocity} V_g\approx 1.07\times10^{-5}c. \ee
Thus the probe filed indeed propagates with a very low group
velocity in $z$ direction which is due to the EIT effect
contributed by the control field.

Note that group-velocity dispersion term (i.e. the term
proportional to $\partial^2 U/\partial \tau^2$) does not appears
in Eq.~(\ref{EQU1}). Thus such equation is valid only for the
probe filed with a large $\tau_0$. To estimate the order of
magnitude of $\tau_0$ for which the dispersion is negligible, we
compare the characteristic dispersion length (defined by $L_{\rm
Disp}={\rm Re}(\tau_{0}^{2}/|\partial^2 K/\partial
\omega^2|_{\omega=0}$) and the diffraction length $L_{\rm Diff}$
defined above. By setting $L_{\rm Disp}=L_{\rm Diff}=1.81$ cm we
obtain $\tau_0= 1.28\times 10^{-6}$ s. Consequently, if $\tau_0$
is much larger than $1.28\times 10^{-6}$ s, $L_{\rm Disp}$ will be
much longer than $L_{\rm Diff}$ and the dispersion effect of the
system can be neglected safely.


\subsection{Slow-light Airy wave packet solutions}\label{IIIB}

We now seek slow-light Airy wave packet solutions of
Eq.~(\ref{EQU1}) for the absence of the SG gradient magnetic
field~\cite{note01}, i.e. $B_1=B_2=0$ and hence $Q=0$.

Since Eq.~(\ref{EQU1}) is a linear one, we can solve it by
taking~\cite{guo}
\be\label{ansatz}
U(\tau,\xi,\eta,s)=\psi(\tau,s)\phi(\tau,\xi,\eta), \ee
with
\be\label{gau}
\psi(\tau,s)=\frac{1}{\sqrt[4]{2\pi\rho^2}}
e^{-(s-\tau/\lambda)^2/(4\rho^2)}=\frac{1}{\sqrt[4]{2\pi\rho^2}}e^{-(z-V_g
t)^2/(4\rho^2L_{\rm Diff}^2)}, \ee
where $\rho$ is a free real parameter. When writing Eq.~(\ref{gau})
we have assumed that the probe-field envelope is a Gaussian pulse
propagating in $z$ direction with velocity $V_g$.

In this way, $\phi(\tau,\xi,\eta)$ satisfies the following equation
\bea\label{phi}
&& i\lambda\frac{\partial}{\partial
\tau}\phi+\frac{1}{2}\left(\frac{\partial^{2} }{\partial
\xi^{2}}+\frac{\partial^{2} }{\partial \eta^{2}}\right)\phi=0.
\eea
Taking $\phi(\tau,\xi,\eta)=\phi_1(\tau,\xi)\phi_2(\tau,\eta)$,
Eq.~(\ref{phi}) can be further decomposed into
\bes\label{phi1} \bea
&& i\lambda\frac{\partial \phi_1}{\partial
\tau}+\frac{1}{2}\frac{\partial^{2} \phi_1
}{\partial \xi^{2}}=0,\\
&& i\lambda\frac{\partial \phi_2}{\partial
\tau}+\frac{1}{2}\frac{\partial^{2} \phi_2 }{\partial \eta^{2}}=0,
\eea\ees
which admit the Airy function solutions $\phi_1(\tau,\xi)={\rm
Ai}[\xi-\tau^2/(4\lambda^2)]e^{i[\xi/2-\tau^2/(12\lambda^2)]\tau/\lambda}$
and $\phi_2(\tau,\eta)={\rm
Ai}[\eta-\tau^2/(4\lambda^2)]e^{i[\eta/2-\tau^2/(12\lambda^2)]\tau/\lambda}$,
respectively. Here ${\rm Ai}$ is the Airy function~\cite{Berry}.
Thus, we have $\phi(\tau,\xi,\eta)={\rm
Ai}[\xi-\tau^2/(4\lambda^2)]{\rm
Ai}[\eta-\tau^2/(4\lambda^2)]e^{i[\xi/2+\eta/2-\tau^2/(6\lambda^2)]\tau/\lambda}$.
The Airy wave packet has the property that its intensity profile
remains invariant but experiences a constant acceleration in both
$x$ and $y$ directions during
propagation~\cite{Berry,greenb,Siviloglou1}.

An ideal Airy wave packet, however, is not square integrable, i.e.
$\int {\rm Ai}^2(x)dx\rightarrow\infty$, which means that it has
infinite energy. The reason comes from the fact that the tail of
the Airy function decays very slowly. Thus, it is not possible to
generate an ideal Airy wave packet experimentally. One suitable
way to solve this problem is to use a finite-energy Airy wave
packet by introducing an additional exponential aperture function,
i.e. by taking the initial condition as the form of
$\phi(0,\xi,\eta)={\rm Ai}(\xi){\rm Ai}(\eta)e^{a_1\xi+a_2\eta}$.
Here $a_j$ ($j=1,\,2$) are positive parameters so as to ensure
containment of the infinite Airy tail. Typically, $a_j\ll1$ so the
resulting profile of the function closely resembles that of the
intended Airy function~\cite{Siviloglou1}. Such finite-energy Airy
wave packets have been recently demonstrated in
experiment~\cite{Siviloglou2}. By directly integrating Eq.
(\ref{phi}) we have
\bea \label{airy} \phi(\tau,\xi,\eta)&=&{\rm
Ai}[\xi-\tau^2/(4\lambda^2)+ia_1\xi]{\rm
Ai}[\eta-\tau^2/(4\lambda^2)+ia_2\eta]e^{i[\xi/2+\eta/2-\tau^2/(6\lambda^2)]\tau/\lambda}\nonumber\\
&& \times
e^{a_1\xi-a_1\xi^2/2+ia_1^2\xi/2}e^{a_2\eta-a_2\eta^2/2+ia_2^2\eta/2}.
\eea
The center of the wave packet (\ref{airy}) moves along the
trajectory $\xi=\eta=\tau^2/(4\lambda^2)$ and hence tend to freely
accelerate during propagation even without any action by external
force.

Consequently, the solution of Eq.~(\ref{EQU1}) without the
external potential reads
\bea\label{ALB}
U(\tau,\xi,\eta,s) &=&
\frac{1}{\sqrt[4]{2\pi\rho^2}}e^{-(s-\tau/\lambda)^2/(4\rho^2)}\, {\rm
Ai}[\xi-\tau^2/(4\lambda^2)+ia_1\xi]{\rm
Ai}[\eta-\tau^2/(4\lambda^2)+ia_2\eta]\nonumber\\
&& \times e^{i[\xi/2+\eta/2-\tau^2/(6\lambda^2)]\tau/\lambda}
e^{a_1\xi-a_1\xi^2/2+ia_1^2\xi/2}e^{a_2\eta-a_2\eta^2/2+ia_2^2\eta/2},
\eea
which consists of two Airy wave packets in $x$ and $y$ directions
and a Gaussian pulse in $z$ direction. The center of the probe filed (\ref{ALB}) moves along with the trajectory
\be
(x(t),y(t),z(t)\,)=\left(\frac{R}{4\lambda^2\tau_0^2}t^2,\frac{R}{4\lambda^2\tau_0^2}t^2,V_g t\right).
\ee
Notice that the solution (\ref{ALB}) is localized in all three
spatial dimensions and in time, thus it can be considered as a
{\it (3+1)D (linear) Airy light bullet} realized in the coherent
atomic system via EIT, which has an ultraslow propagating velocity
in $z$ direction.

The Airy light wave packet obtained above is quite stable.
Fig.~\ref{fig3} shows the result of numerical simulation on its
%
\begin{figure}[tbph]
\centering
\includegraphics[scale=0.75]{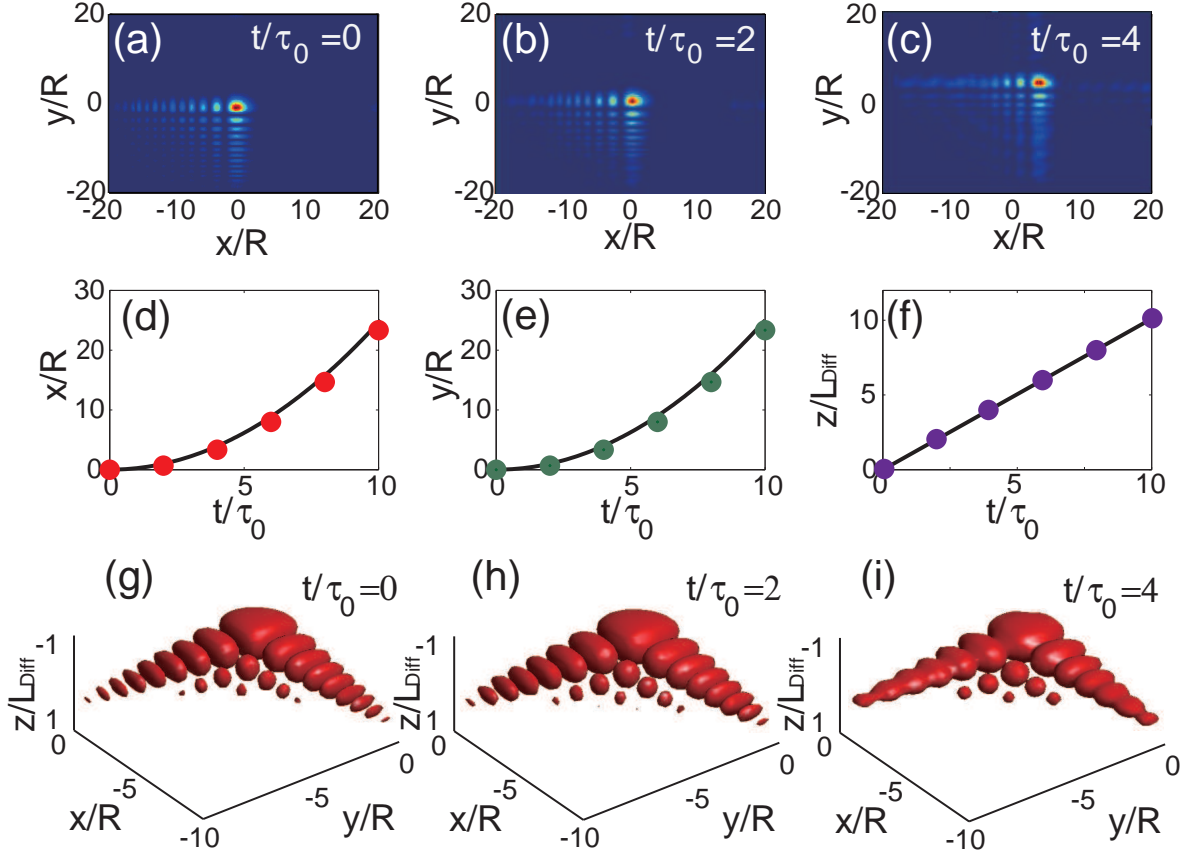}
\caption{{\protect\footnotesize (Color online) (a)-(c) Spatial
intensity distributions of the slow-light Airy wave packet in $xy$
plane for $t/\tau_0=0$, $2$, and $4$, respectively. (d)-(f)
Numerical (dots) and theoretical (solid lines) results of the
center positions $x=x(t)$, $y=y(t)$, $z=z(t)$ of the slow-light
Airy wave packet. (g)-(i) Intensity isosurfaces of the Airy wave
packet for $t/\tau_0=0$, $2$, and $4$, respectively. }}
\label{fig3}
\end{figure}
%
stability by using split-step Fourier method. In doing this, we
have added small random perturbations less than 10\% to both
amplitude and phase to the solution Eq.~(\ref{ALB}) and then
evolve it according to Eq.~(\ref{EQU1}) with $Q=0$. The values of
parameters are given in Sec.~\ref{IIIA} with $\tau_0=L_{\rm
Diff}/V_g=0.56\times10^{-5}$ s (i.e. $\lambda=1$).
Fig.~\ref{fig3}(a)-(c) show the spatial distribution of the Airy
wave packet in the $xy$ plane for $t/\tau_0=0$, $2$, and $4$,
respectively. Fig.~\ref{fig3}(d)-(f) show the numerical (dots) and
theoretical (lines) shifts of the Airy wave packet in the $x$,
$y$, and $z$ directions. The accelerating behavior of wave packets
in $x$ and $y$ directions is clearly observed. In
Fig.~\ref{fig3}(g)-(i) we show the intensity isosurfaces of the
Airy wave packet for $t/\tau_0=0$, $2$, and $4$ in $xyz$ space,
respectively. These results clearly demonstrate that the
slow-light Airy wave packet obtained in the present system is
rather robust up to the propagation time
$t=4\tau_0\approx2.24\times10^{-5}$ s even without any trapping
potential. This stability can be explained by the low absorption
of the system  and the capability of Airy wave packets in $x$ and
$y$ directions for withstanding diffraction.

\subsection{Acceleration control of the slow-light Airy wave packets}\label{IIIC}

In the paper by Berry and Balazs~\cite{Berry}, motion of an Airy
wave packet obeying a (1+1)D Schr\"{o}dinger equation with a time
dependent but spatially uniform force was investigated. Here we
extend their study to (3+1)D case and realize an active control of
the acceleration by using the SG gradient magnetic field.

The explicit form of the potential in Eq.~(\ref{EQU1}) reads
\be\label{potential}
Q(\xi,\eta)=\kappa_{13}\frac{|\Omega_{c}|^2\mu_{21}
+d_{21}^{(0)\,2}\mu_{31}}{(|\Omega_{c}|^{2}-d_{21}^{(0)}d_{31}^{(0)})^2}L_{\rm
Disp}R(B_1\xi+B_2\eta)\equiv Q_1\xi+Q_2\eta. \ee
We see that the coefficients $Q_1$ and $Q_2$ are proportional to
the gradient of the SG magnetic field in $x$ and $y$ directions,
respectively. With such potential, Eq.~(\ref{phi1}) are replaced by
\bes\label{phi2}\bea
&& i\lambda\frac{\partial \phi_1}{\partial \tau}+\frac{1}{2}\frac{\partial^{2} \phi_1
}{\partial \xi^{2}}+Q_1\xi\phi_1=0,\\
&& i\lambda\frac{\partial \phi_2}{\partial
\tau}+\frac{1}{2}\frac{\partial^{2} \phi_2 }{\partial
\eta^{2}}+Q_2\eta\phi_2=0.
\eea\ees
Using the transformation
$\phi_1=\phi_1'e^{i[Q_1\xi'+Q_1^2\tau^2/(3\lambda^2)]\tau/\lambda}$ and
$\phi_2=\phi_2'e^{i[Q_2\eta'+Q_2^2\tau^2/(3\lambda^2)]\tau/\lambda}$ with
$\xi'=\xi-Q_1\tau^2/(2\lambda^2)$ and $\eta'=\eta-Q_2\tau^2/(2\lambda^2)$,
Eq.~(\ref{phi2}) is converted into $i\lambda \partial \phi_1'/\partial
\tau+(1/2)\partial^2\phi_1'/\partial \xi'^2=0$ and $i\lambda \partial
\phi_2'/\partial \tau+(1/2)\partial^2\phi_2'/\partial \eta'^2=0$.
Thus we can obtain the Airy function solutions of
Eq.~({\ref{phi2}) as  $\phi_1(\tau,\xi')={\rm
Ai}[\xi'-\tau^2/(4\lambda^2)]e^{i[\xi'/2-\tau^2/(12\lambda^2)]\tau/\lambda}e^{i[Q_1\xi'+Q_1^2\tau^2/(3\lambda^2)]\tau/\lambda}$
and $\phi_2(\tau,\eta')={\rm
Ai}[\eta'-\tau^2/(4\lambda^2)]e^{i[\eta'/2-\tau^2/(12\lambda^2)]\tau/\lambda}e^{i[Q_2\eta'+Q_2^2\tau^2/(3\lambda^2)]\tau/\lambda}$, and hence $\phi(\tau,\xi',\eta')=\phi_1(\tau,\xi')\phi_2(\tau,\eta')={\rm
Ai}[\xi'-\tau^2/(4\lambda^2)]{\rm
Ai}[\eta'-\tau^2/(4\lambda^2)]e^{i[\xi'/2+\eta'/2-\tau^2/(6\lambda^2)]\tau/\lambda}
e^{i[Q_1\xi'+Q_2\eta'+Q_1^2\tau^2/(3\lambda^2)+Q_2^2\tau^2/(3\lambda^2)]\tau/\lambda}$.
A finite-energy Airy function solution reads
\bea \label{airy1} \phi(\tau,\xi',\eta')&=&{\rm
Ai}[\xi'-\tau^2/(4\lambda^2)+ia_1\xi']{\rm
Ai}[\eta'-\tau^2/(4\lambda^2)+ia_2\eta']e^{i[\xi'/2+\eta'/2-\tau^2/(6\lambda^2)]\tau/\lambda}\nonumber\\
&& \times
e^{a_1\xi'-a_1\xi'^2/2+ia_1^2\xi'/2}e^{a_2\eta'-a_2\eta'^2/2
+ia_2^2\eta'/2}e^{i[Q_1\xi'+Q_2\eta'+Q_1^2\tau^2/(3\lambda^2)+Q_2^2\tau^2/(3\lambda^2)]\tau/\lambda}.
\eea
Consequently, the Airy light wave packet solution of
Eq.~(\ref{EQU1}) with the potential (\ref{potential}) is given by
\bea\label{ALB1}
U(\tau,\xi',\eta',s) &=& \frac{1}{\sqrt[4]{2\pi\rho^2}}e^{-(s-\tau/\lambda)^2/(4\rho^2)}\,{\rm
Ai}[\xi'-\tau^2/(4\lambda^2)+ia_1\xi']{\rm
Ai}[\eta'-\tau^2/(4\lambda^2)+ia_2\eta']\nonumber\\
&& \times e^{i[\xi'/2+\eta'/2-\tau^2/(6\lambda^2)]\tau/\lambda}
e^{a_1\xi'-a_1\xi'^2/2+ia_1^2\xi'/2}e^{a_2\eta'-a_2\eta'^2/2
+ia_2^2\eta'/2}\nonumber\\
&& \times e^{i[Q_1\xi'+Q_2\eta'+Q_1^2\tau^2/(3\lambda^2)+Q_2^2\tau^2/(3\lambda^2)]\tau/\lambda}.
\eea
It is clear that the motional trajectory of the Airy wave packet
(\ref{ALB1}) in the presence of the SG gradient magnetic field is
given by $\xi=[\tau^2/(2\lambda^2)](1/2+Q_1)$,
$\eta=[\tau^2/(2\lambda^2)](1/2+Q_2)$, and $s=\tau/\lambda$, i.e.
\be
(x(t),y(t),z(t)\,)=\left(\frac{R}{2\lambda^2\tau_0^2}\left(\frac{1}{2}
+Q_1\right)t^2,\frac{R}{2\lambda^2\tau_0^2}\left(\frac{1}{2}+Q_2\right)t^2,V_g t\right). \ee

Obviously, the magnitude and direction of the acceleration of the
slow-light Airy wave packet in the $xy$ plane can be easily
controlled by changing the gradient of the SG magnetic field, i.e.
by changing the magnitude of $Q_1$ and $Q_2$. Particularly, when
taking $Q_1=Q_2=-1/2$ corresponding to the magnetic gradients
$B_1=B_2\approx-3.19$ G cm$^{-1}$ (the minus sign here means the
SG magnetic field should be applied along $-z$ direction), the
trajectory of the Airy wave packet becomes $(\xi,\eta)=(0,0)$.
This means that the force provided by the SG magnetic field in
each transverse direction is sufficient to overcome the
acceleration in this direction, i.e. the transverse motion of the
wave packet can be completely stopped. As a result, the Airy wave
packet moves along $z$ direction with the ultraslow velocity
$V_g$.

Additional control of the transverse motion of the Airy wave
packet is also possible. In fact, if the SG gradient magnetic
field is chosen to be time-dependent, i.e. $Q_1=Q_1(\tau)$ and
$Q_2=Q_2(\tau)$, the transverse trajectory of the Airy wave packet
becomes
\bes\bea
&& \xi=\frac{1}{\lambda^2}\left[\frac{\tau^2}{4}+\int_0^{\tau} Q_1(\tau')(\tau-\tau')d\tau'\right], \\
&& \eta=\frac{1}{\lambda^2}\left[\frac{\tau^2}{4}+\int_0^{\tau}
Q_2(\tau')(\tau-\tau')d\tau'\right]. \eea\ees
Particularly, when taking $Q_1=-1/2+V_1\delta(\tau)$ and
$Q_2=-1/2$, the trajectory of the Airy wave packet is given by
$(\xi,\eta,s)=(V_1\tau/\lambda^2,0,\tau/\lambda)$. This means that
the wave packet propagates with a constant velocity
$V_1R/(\lambda^2\tau_0)$ in the $x$ direction and locates at zero
in the $y$ direction. Furthermore, if taking
$Q_1=-1/2+V_1\delta(\tau)$ and $Q_2=-1/2+V_2\delta(\tau)$, the
trajectory of the wave packet turns to
$(\xi,\eta,s)=(V_1\tau/\lambda^2,V_2\tau/\lambda^2,\tau/\lambda)$,
i.e. it propagates with constant velocities
$V_1R/(\lambda^2\tau_0)$ and $V_2R/(\lambda^2\tau_0)$ in $x$ and
$y$ directions, respectively.

Shown in Fig.~\ref{fig4} is the trajectory control of the
slow-light Airy wave packets by changing the values of $V_1$ and
$V_2$. The values of parameters are the same with those used in
Fig.~\ref{fig3}. In Fig.~\ref{fig4}(a)-(c) we show spatial
distributions of the Airy wave packet in the $xy$ plane at
$t/\tau_0=4$ for $(V_1,V_2)=(0,0)$, $(V_1,V_2)=(1,0)$, and
$(V_1,V_2)=(1,1)$, respectively. Fig.~\ref{fig4}(c) shows the
corresponding 3D trajectory plots of Airy wave packets
respectively for $(V_1,V_2)=(0,0)$ (dotted line),
$(V_1,V_2)=(1,0)$ (dashed line), and $(V_1,V_2)=(1,1)$ (solid
line) with the fixed atomic medium length $4L_{\rm
Diff}\approx7.24$ cm. The center positions of Airy wave packets
when exiting the medium are $(0,0)$, $(4,0)$, and $(4,4)$,
respectively.
%
\begin{figure}[tbph]
\centering
\includegraphics[scale=0.75]{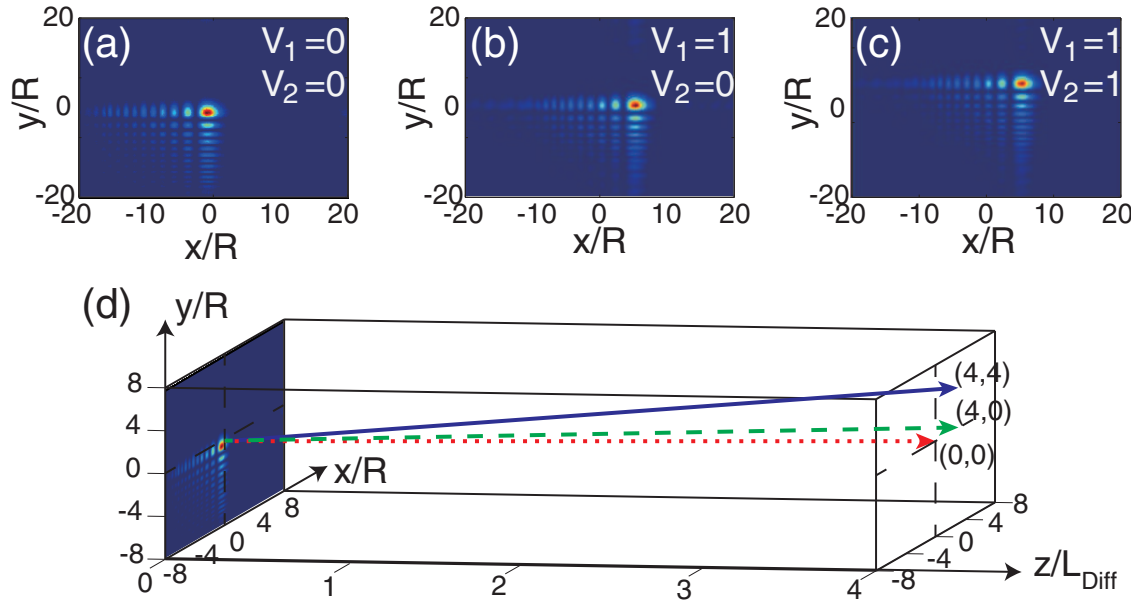}
\caption{{\protect\footnotesize (Color online) (a)-(c): Spatial
intensity distributions of the slow-light Airy wave packet in $xy$
plane at $t/\tau_0=4$ for $(V_1,V_2)=(0,0)$, $(V_1,V_2)=(1,0)$,
and $(V_1,V_2)=(1,1)$, respectively. (d): The 3D trajectory plots
of Airy wave packets for $(V_1,V_2)=(0,0)$ (dotted line),
$(V_1,V_2)=(1,0)$ (dashed line), and $(V_1,V_2)=(1,1)$ (solid
line). The medium length is fixed to be $4L_{\rm Diff}\approx7.24$
cm. The center positions of Airy wave packets when exiting the
medium are $(0,0)$, $(4,0)$, and $(4,4)$, respectively.}}
\label{fig4}
\end{figure}
%
One can also obtain easily other different motional trajectories
of the slow-light Airy wave packet by using different SG gradient
magnetic field, which are not shown here.


\section{Slow-light Airy wave packets in the presence of dispersion}\label{IV}


The results presented in the last section are valid only for large
time duration of the probe filed, i.e for large $\tau_0$. If $\tau_0$
becomes smaller, the dispersion effect of the system is significant and
hence must be taken into account. In this case, the envelope equation
(\ref{EQU1}) is not valid. To get an envelope equation valid for
the presence of dispersion, the multi-scale variables
used in Sec.~\ref{IIIA} should be replaced by
$z_{j}=\epsilon^{j/2} z$ ($j=0,1,2$), $t_{j}=\epsilon^{j/2} t$
($j=0,1$), $x_{1}=\epsilon^{1/2} x$, and $y_{1}=\epsilon^{1/2} y$.
Then, at the leading order ($l=1$, i.e. the terms of order of
$\epsilon$), we have the same solutions with those given in Eqs.
(\ref{ord1}) and (\ref{Disp}) in Sec \ref{IIIA}. At next order
($l=2$, i.e. the terms of order of $\epsilon^{3/2}$), a
divergence-free condition requires $\partial F/\partial
z_1+(1/V_g)\partial F/\partial t_1=0$. The divergence-free
condition at the third order ($l=3$, i.e. the terms of order of
$\epsilon^2$) yields the equation for the envelope function $F$:
\be  i\frac{\partial }{\partial
z_{2}}F-\frac{1}{2}\frac{\partial^2 K}{\partial
\omega^2}\frac{\partial^{2} }{\partial t_1^{2}}F+\frac{c}{2\omega
_{p}}\left( \frac{\partial ^{2}}{\partial
x_{1}^{2}}+\frac{\partial ^{2}}{\partial y_{1}^{2}}\right)F+
P(x_1,y_1)F=0. \ee

Combining the equations of all orders and returning to original
variables, we obtain the equation in the dimensionless form
\be\label{EQU2} i\frac{\partial }{\partial
s}U+\frac{1}{2}\left(\mu\frac{\partial^{2} }{\partial
\sigma^{2}}+\frac{\partial^{2} }{\partial
\xi^{2}}+\frac{\partial^{2}}{\partial \eta^{2}}\right)U
+Q(\xi,\eta)U=0, \ee
with $\sigma=(t-z/V_g)/\tau_0$ and $\mu=-{\rm sign}[{\rm Re}
(\partial^2 K/\partial \omega^2|_{\omega=0})]L_{\rm Diff}/L_{\rm
Disp}$. The quantities $s$, $\xi$, $\eta$, $U$, and $Q$ have the same definitions as
those in given Eq.~(\ref{EQU1}).

With the parameters given in Sec.~\ref{IIIA} and by taking
$\tau_0=1.28 \times 10^{-6}$ s (which is the critical value for
which the dispersion effect must be considered; see the last
paragraph of Sec.~\ref{IIIA}), we obtain
$K_2=(-0.87+i0.26)\times10^{-12}$ cm$^{-1}$ s$^2$ which leads to
$L_{\rm Disp}=L_{\rm Diff}=1.81$ cm, and hence $\mu=1$.

In the absence of the SG gradient magnetic field (i.e. $Q=0$), by
using a similar method in Sec.~\ref{IIIB} we obtain the Airy wave
packet solution of Eq.~(\ref{EQU2})
%
\bea\label{Airy3}
U(\sigma,\xi,\eta,s) &=& {\rm Ai}(\sigma-s^2/4+ia_0\sigma)\,{\rm
Ai}(\xi-s^2/4+ia_1\xi)\,{\rm
     Ai}(\eta-s^2/4+ia_2\eta) \nonumber\\
 & & \times e^{i(\sigma/2+\xi/2+\eta/2-s^2/4)s}
     e^{a_0\sigma-a_0\sigma^2/2+ia_0^2
     \sigma/2}e^{a_1\xi-a_1\xi^2/2+ia_1^2\xi/2}\nonumber\\
 & & \times e^{a_2\eta-a_2\eta^2/2+ia_2^2\eta/2},
\eea
which consists of two Airy beams in $x$ and $y$ directions and a
longitudinal Airy wave packet propagating in $z$ direction
\cite{note2}. Different from the solution (\ref{ALB}), the center
of the Airy wave packet given by (\ref{Airy3}) moves along the
trajectory $\xi=\eta=\sigma=s^2/4$. That is, in $x$ and $y$
directions
\be \label{traj1}
(x(z),y(z)\,)=\left(\frac{R}{4L_{\rm
Diff}^2}z^2,\frac{R}{4L_{\rm Diff}^2}z^2\right),
\ee
and in $z$ direction
\be \label{traj2} z(t)=2L_{\rm
Diff}\left(\sqrt{\frac{t}{\tau_0}+\lambda^2}-\lambda\right), \ee
where $\lambda$ is defined by Eq. (\ref{lambda}). Thus the
propagating velocity of the Airy wave packet in $z$ direction is
\be \label{groupvelocity1} V_z=\frac{V_g}{z/(2\lambda L_{\rm Diff})+1}. \ee
From Eq.~(\ref{traj1}), we see that the Airy light wave packet
(\ref{Airy3}) has stationary, but bent beam intensity
distributions in the transverse $x$ and $y$ directions; in the
longitudinal $z$ direction it however is a Airy spatial-temporal
wave packet with the propagating velocity $V_z$, which is
proportional to $V_g$ (Eq.~(\ref{groupvelocity1})\,).
Interestingly, $V_z$ can be further reduced when the propagating
distance $z$ becomes large. For instance, with the given
parameters we have $\lambda=4.38$, and hence we obtain
$V_z=0.68V_g$ when $z=4L_{\rm Diff}$.

Shown in Fig.~\ref{fig5} is the result of a numerical simulation
of the slow-light Airy wave packet in the presence of dispersion.
In the simulation,
%
\begin{figure}[tbph]
\centering
\includegraphics[scale=0.75]{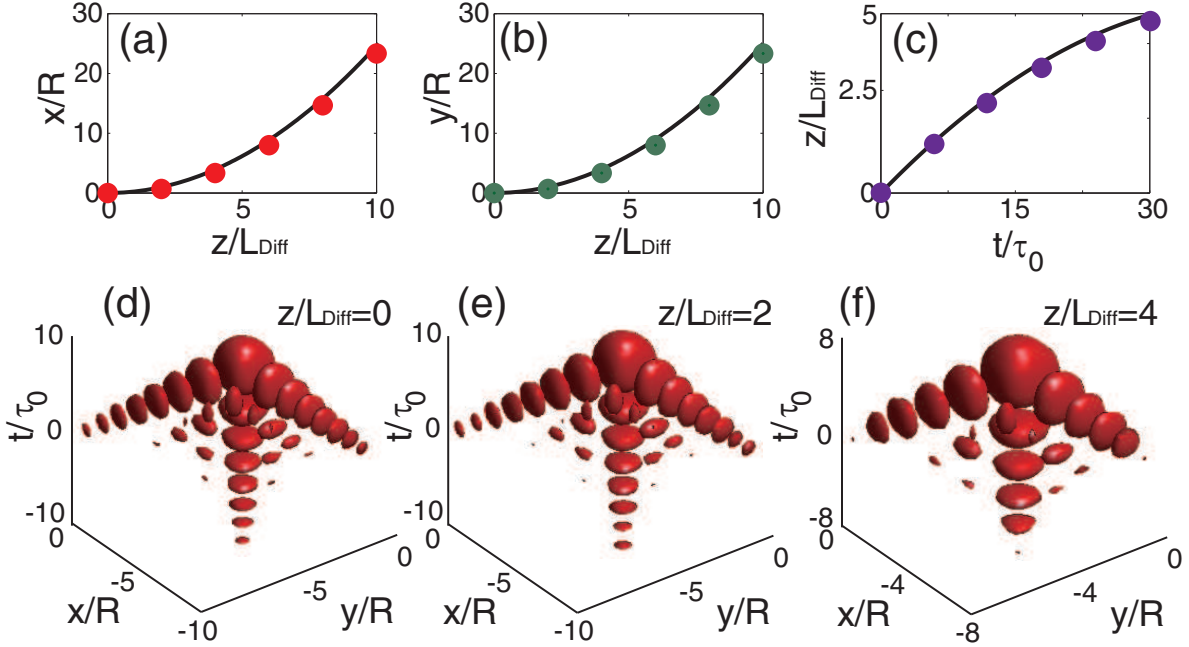}
\caption{{\protect\footnotesize (Color online) (a)-(c) The
numerical (dots) and theoretical (lines) results of the center
positions $x=x(z)$, $y=y(z)$, $z=z(t)$ of the show-light Airy wave
packet in the presence of dispersion. (d)-(f) Intensity
isosurfaces of the show-light Airy wave packet for $z/L_{\rm
Diff}=0$, $2$, and $4$, respectively.}} \label{fig5}
\end{figure}
%
to test the stability of the Airy wave packet we have added small
random perturbations less than 10\% to both amplitude and phase to
the solution Eq.~(\ref{Airy3}) and then evolve it according to
Eq.~(\ref{EQU2}) with $Q=0$. Plotted in Fig.~\ref{fig5}(a)-(c) are
the numerical (dots) and theoretical (lines) results on the
evolution of the center position $x=x(z)$, $y=y(z)$, $z=z(t)$ of
the Airy wave packet. The bending (in $x$ and $y$ directions) and
decelerating (in $z$ direction) behavior is clearly observed. In
Fig.~\ref{fig5}(d)-(f) we show the intensity isosurfaces of the
Airy wave packet for $z/L_{\rm Diff}=0$, $2$, and $4$ in $xyt$
space, respectively. These results demonstrate that the slow-light
Airy wave packet obtained in the present system is rather robust
up to the propagation distance $z=4L_{\rm Diff}\approx7.24$ cm
even without any trapping potential. This stability can be
explained by the low absorption and the capability of the Airy
wave packet for withstanding both dispersion and diffraction.

In the presence of the SG magnetic field, the slow-light Airy wave
packet solution of Eq.~(\ref{EQU2}) with the potential
(\ref{potential}) can also be obtained, which reads
\bea\label{Airy31}
U(\sigma,\xi',\eta',s) &=& {\rm Ai}(\sigma-s^2/4+ia_0\sigma){\rm
Ai}(\xi'-s^2/4+ia_1\xi'){\rm
Ai}(\eta'-s^2/4+ia_2\eta')\nonumber\\
&& \times e^{i(\sigma/2+\xi'/2+\eta'/2-s^2/4)s}
e^{a_0\sigma-a_0\sigma^2/2+ia_0^2\sigma/2}e^{a_1\xi'
-a_1\xi'^2/2+ia_1^2\xi'/2}e^{a_2\eta'-a_2\eta'^2/2
+ia_2^2\eta'/2}\nonumber\\
&& \times e^{i(Q_1\xi'+Q_2\eta'+Q_1^2s^2/3+Q_2^2s^2/3)s}, \eea
where $\xi'=\xi-Q_1s^2/2$ and $\eta'=\eta-Q_2s^2/2$. We see that,
as in the case without the SG magnetic field (see
Eq.~(\ref{Airy3})\,), the Airy wave packet propagates in $z$
direction with the same ultraslow, decreased velocity $V_z$ given
in Eq.~(\ref{groupvelocity1}). In addition, it has also a
stationary intensity distribution bent in $x$ and $y$ directions,
but now with a different bending trajectory
$\xi=(s^2/2)(1/2+Q_1)$, $\eta=(s^2/2)(1/2+Q_2)$, i.e.
\be (x(z),y(z)\,)=\left(\frac{R}{2L_{\rm
Diff}^2}\left(\frac{1}{2}+Q_1\right)z^2,\frac{R}{2L_{\rm
Diff}^2}\left(\frac{1}{2}+Q_2\right)z^2\,\right). \ee
Obviously, the trajectory bending in the transverse directions of
the Airy wave packet can be completely eliminated by using the SG
gradient magnetic field. For example, taking $Q_1=Q_2=-1/2$ one
has $(x(z),y(z)\,)=(0,0)$. Similarly, other kinds of active
control on the trajectory of Airy wave packets can also be
implemented easily by manipulating the SG gradient magnetic field.

\section{Summary}

In this article, we have proposed a scheme to create
(3+1)-dimensional slow-light Airy wave packets in a resonant
$\Lambda$-type three-level atomic gas via EIT induced by the
control field. We have shown that in the absence of dispersion the
Airy wave packets obtained consist of two Airy wave packets
accelerated in transverse directions and a longitudinal Gaussian
pulse with a constant propagating velocity lowered to
$10^{-5}\,c$. We have also shown that in the presence of
dispersion one is able to create another type of slow-light Airy
wave packets consisting of two Airy beams in transverse directions
and an Airy wave packet in the longitudinal direction. In this
situation, the longitudinal velocity of the Airy wave packets can
be further reduced during propagation. In addition, we have
demonstrated that the transverse accelerations (or bending) of the
both types of slow-light Airy wave packets can be completely
eliminated and the motional trajectories of them can be actively
manipulated and controlled by using a Stern-Gerlach gradient
magnetic field. The research presented here opens an avenue for
the exploration of magneto-optical control on Airy beams and wave
packets, and the results obtained in this work may guide new
experimental findings of slow-light Airy wave packets and have
potential applications in optical information processing and
transmission.


\acknowledgments This work was supported by the NSF-China under
Grant Numbers 11174080 and 11105052, and by the Open Fund from the
State Key Laboratory of Precision Spectroscopy, ECNU.




\begin{references}
\bibitem{note001}Here beam  means that the intensity
distribution of the light field is stationary, wave packet (pulse)
means that the intensity is localized in both space and time with
larger (smaller) spatial-temporal width.
\bibitem{note01}In our model, the potential contributed by the SG gradient magnetic field is not for stability but for providing an acceleration control of the Airy light bullets.
\bibitem{note1}The frequency and wave number of the probe field are given by
$\omega_p+\omega$ and $k_p+K(\omega),$ respectively. Thus
$\omega=0$ corresponds to the center frequency of the probe
field.
\end{references}
\end{document}